\documentclass[twocolumn,showpacs,preprintnumbers,amsmath,amssymb]{revtex4}

\usepackage{graphicx}
\usepackage{dcolumn}
\usepackage{bm}

\begin{document}

\preprint{APS/123-QED}

\title{Te 5$p$ orbitals bring three-dimensional electronic structure to two-dimensional Ir$_{0.95}$Pt$_{0.05}$Te$_2$}

\author{D.~Ootsuki$^1$}
\author{T.~Toriyama$^2$}
\author{M.~Kobayashi$^3$}
\author{S.~Pyon$^3$}
\author{K.~Kudo$^3$}
\author{M.~Nohara$^3$}
\author{K.~Horiba$^4$}
\author{M.~Kobayashi$^4$}
\author{K.~Ono$^4$}
\author{H.~Kumigashira$^4$}
\author{T.~Noda$^1$}
\author{T.~Sugimoto$^1$}
\author{A.~Fujimori$^{5}$}
\author{N.~L.~Saini$^{6}$}
\author{T.~Konishi$^7$}
\author{Y.~Ohta$^2$}
\author{T.~Mizokawa$^{1}$}

\affiliation{$^1$Department of Physics and Department of Complexity Science and Engineering, 
University of Tokyo, 5-1-5 Kashiwanoha, Chiba 277-8561, Japan}
\affiliation{$^2$Department of Physics, Chiba University, Inage-ku, Chiba 263-8522, Japan}
\affiliation{$^3$Department of Physics, Okayama University, Kita-ku, Okayama 700-8530, Japan}
\affiliation{$^4$Institute of Materials Structure Science, High Energy Accelerator Research Organization (KEK), 
                 Tsukuba, Ibaraki 305-0801, Japan}
\affiliation{$^5$Department of Physics, University of Tokyo, 7-3-1 Hongo, Tokyo 113-0033, Japan}
\affiliation{$^6$Department of Physics, University of Roma "La Sapienza" Piazalle Aldo Moro 2, 00185 Roma, Italy}
\affiliation{$^7$Graduate School of Advanced Integration Science, Chiba University, Chiba 263-8522, Japan}

\date{\today}

\begin{abstract}
We have studied the nature of the three-dimensional multi-band electronic structure 
in the two-dimensional triangular lattice Ir$_{1-x}$Pt$_{x}$Te$_2$ ($x$=0.05) superconductor 
using angle-resolved photoemission spectroscopy (ARPES), x-ray photoemission spectroscopy (XPS)
and band structure calculation. ARPES results clearly show a cylindrical (almost two-dimensional) 
Fermi surface around the zone center. Near the zone boundary, the cylindrical Fermi surface is 
truncated into several pieces in a complicated manner with strong three-dimensionality. 
The XPS result and the band structure calculation indicate that the strong Te 5$p$-Te 5$p$ 
hybridization between the IrTe$_2$ triangular lattice layers is responsible for 
the three-dimensionality of the Fermi surfaces and the intervening of the Fermi surfaces
observed by ARPES.
\end{abstract}

\maketitle

$3d$, $4d$, and $5d$ transition-metal compounds, and 5$f$ actinide compounds
with layered crystal structures often exhibit three-dimensional multi-band 
Fermi surfaces, which can induce complicated spin-charge-orbital instabilities
due to interplay between the two-dimensional electronic structure of the layer
and the interaction between neighboring layers. 
The layered compounds tend to have flat cleavage surfaces which are suitable 
for angle-resolved photoemission spectroscopy (ARPES) measurements, and
the three-dimensional electronic structures can be observed
by sweeping photon energy for ARPES.
For example, the ARPES study on the classical 1T-TaSe$_2$ system 
has shown that the large Se-Se interaction between 
the layers and the strong covalency between the Ta 5$d$ and Se 4$p$ orbitals 
provide three-dimensional Fermi surfaces which play important roles 
in the charge density wave formation \cite{Horiba2002}.
In the case of Fe-based superconductors, the origin of the spin and orbital order
in the parent materials is still controversial and the three-dimensional 
multi-band Fermi surfaces observed by ARPES are key ingredients to understand 
the spin and orbital instabilities as well as the nodeless and nodal superconducting 
states \cite{Yoshida2011}.
In addition, the recent ARPES study on URu$_2$Si$_2$ (which has the ThCr$_2$Si$_2$ structure) 
has revealed that the nesting vectors in the three-dimensional momentum space are associated with the
"hidden" order \cite{Meng2013}.

Very recently, the layered $5d$ transition-metal chalcogenide IrTe$_2$
has been attracting great interest due to the discovery of superconductivity 
in doped or intercalated IrTe$_2$ by Pyon {\it et al.} \cite{Pyon2012} 
and by Yang {\it et al.} \cite{Yang2012}.
IrTe$_2$ undergoes a structural phase transition at $\sim$ 270 K 
from the trigonal (P3m-1) to the monoclinic (C2/m) structure
\cite{Matsumoto1999}. 
Since the Ir 5$d$-to-Te 5$p$ charge-transfer energy is found 
to be small \cite{Ootsuki2012}, the strong Ir 5$d$-Te 5$p$ hybridization 
with the IrTe$_2$ layer and the Te 5$p$-Te 5$p$ hybridization between
the IrTe$_2$ layers can play significant roles in the structural phase transition
as proposed by Fang {\it et al.} and Oh {\it et al.} \cite{Fang2012,Oh2013}.
In addition, band-structure calculations predict that the multi-band Fermi
surfaces of undistorted and distorted IrTe$_2$ are three dimensional
in spite of its layered structure \cite{Yang2012,Fang2012}. 
Although the multi-band electronic structure of the Ir 5$d$ and Te 5$p$ orbitals
in IrTe$_2$ has been studied using ARPES \cite{Ootsuki2013}, the three dimensionality
of the Fermi surfaces has not been observed in IrTe$_2$ and its relatives.
Here, fundamental questions to be addressed are (i) whether the triangular lattice 
Ir$_{1-x}$Pt$_x$Te$_2$ has the three dimensional Fermi surfaces as predicted by the band-structure calculation
and (ii) what are the characteristics of the observed band structures in the triangular lattice Ir$_{1-x}$Pt$_x$Te$_2$.
In the present study, we have performed ARPES measurements 
of Ir$_{0.95}$Pt$_{0.05}$Te$_2$ at various photon energies
in order to study the three dimensional Fermi surfaces of 
the undistorted triangular lattice superconductor Ir$_{1-x}$Pt$_x$Te$_2$.

The single crystal samples of Ir$_{0.95}$Pt$_{0.05}$Te$_2$ 
were prepared as reported in the literature \cite{Pyon2013}.
The ARPES measurements were performed at beam line 28A 
of Photon Factory, KEK using a SCIENTA SES-2002 electron analyzer 
with circularly polarized light. 
The total energy resolution was set to 20-30 meV for 
the excitation energies from $h\nu=79$ eV to $h\nu=54$ eV. 
The base pressure of the spectrometer was in the $10^{-9}$ Pa range. 
The single crystals of Ir$_{0.95}$Pt$_{0.05}$Te$_2$,
which were oriented by {\it ex situ} Laue measurements, 
were cleaved at 20 K under the ultrahigh vacuum and 
the spectra were acquired within 24 hours after the cleaving. 
The x-ray photoemission spectroscopy (XPS) measurement was 
carried out using JEOL JPS9200 analyzer. 
Mg K$\alpha$ (1253.6 eV) was used as x-ray source. 
The total energy resolution was set to $\sim$ 1.0 eV, and 
the binding energy was calibrated using the Au 4$f$ core level 
of the gold reference sample. 
For the band structure calculations, we employ the code WIEN2k \cite{Blaha2002} 
based on the full-potential linearized augmented-plane-wave method 
and present the calculated results obtained in the generalized gradient 
approximation (GGA) for electron correlations, where we use the 
exchange-correlation potential of Ref. \cite{Perdew1996}.  
The spin-orbit interaction is taken into account for both Ir and Te ions.

\begin{figure}
\includegraphics[width=9cm]{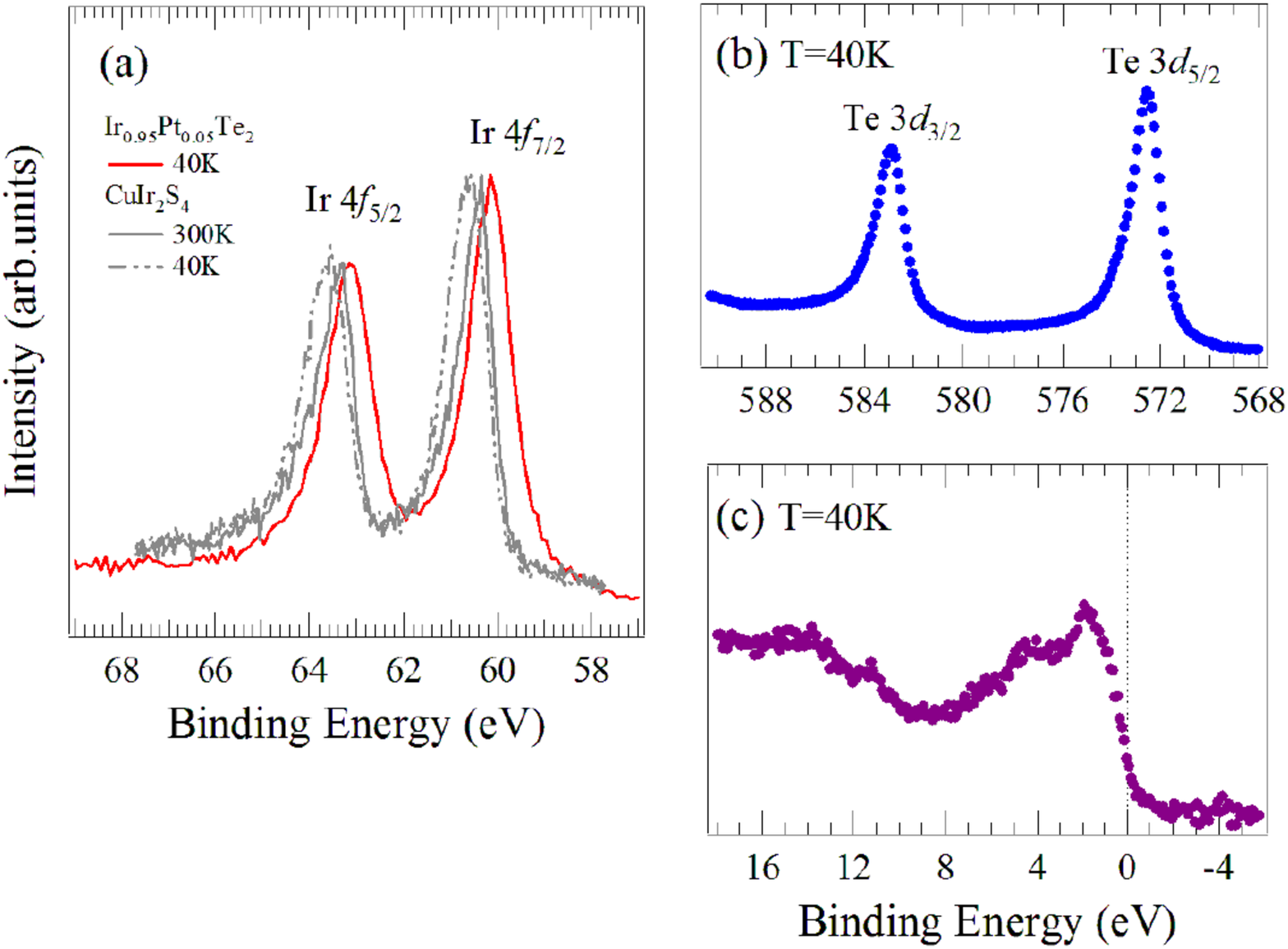}
\caption{(color online)
XPS spectra of Ir$_{0.95}$Pt$_{0.05}$Te$_2$ for the Ir 4$f$ core level (a), 
Te 3$d$ core level (b), and valence band (c).
}
\end{figure}

Before presenting the ARPES result, let us discuss the fundamental electronic 
structure of Ir$_{0.95}$Pt$_{0.05}$Te$_2$, which provides the three-dimensional
Fermi surfaces in the layered triangular lattice material, based on the XPS result.
In Fig. 1(a), the Ir 4$f$ core-level XPS spectrum of Ir$_{0.95}$Pt$_{0.05}$Te$_2$ is 
compared with that of CuIr$_2$S$_4$. 
While the formal valence of Ir is +4 in Ir$_{0.95}$Pt$_{0.05}$Te$_2$,
the binding energy of the Ir 4$f$ core level is much lower than that of CuIr$_2$S$_4$
with Ir$^{3.5+}$, indicating that the actual valence of Ir should be smaller than +3.5.
Therefore, some of the Te 5$p$ electrons are transferred from the Ir 5$d$ orbitals, 
and the Te 5$p$ orbitals are expected to be partially unoccupied. 
This picture is consistent with the band structure calculation in which
the electronic states near the Fermi level ($E_F$) have substantial Te 5$p$ character.
In addition, the binding energy of the Te 3$d$ core level [Fig. 1(b)] is relatively 
small compared to other transition-metal tellurides, indicating the Te 5$p$ states 
are heavily involved in the near-$E_F$ electronic states. 
Indeed, as shown in Fig. 1(c), the valence-band XPS spectra 
of Ir$_{0.95}$Pt$_{0.05}$Te$_2$ exhibits a peak around 2.5 eV below $E_F$ 
whereas the spectral weight near $E_F$ is rather weak.
Since the photo-ionization cross section of Ir 5$d$ is much larger than
that of Te 5$p$ at this photon energy, the valence-band XPS result suggests 
that the electronic states 2-3 eV below $E_F$ are dominated by Ir 5$d$
while those near $E_F$ have more Te 5$p$ character.

\begin{figure}
\includegraphics[width=9cm]{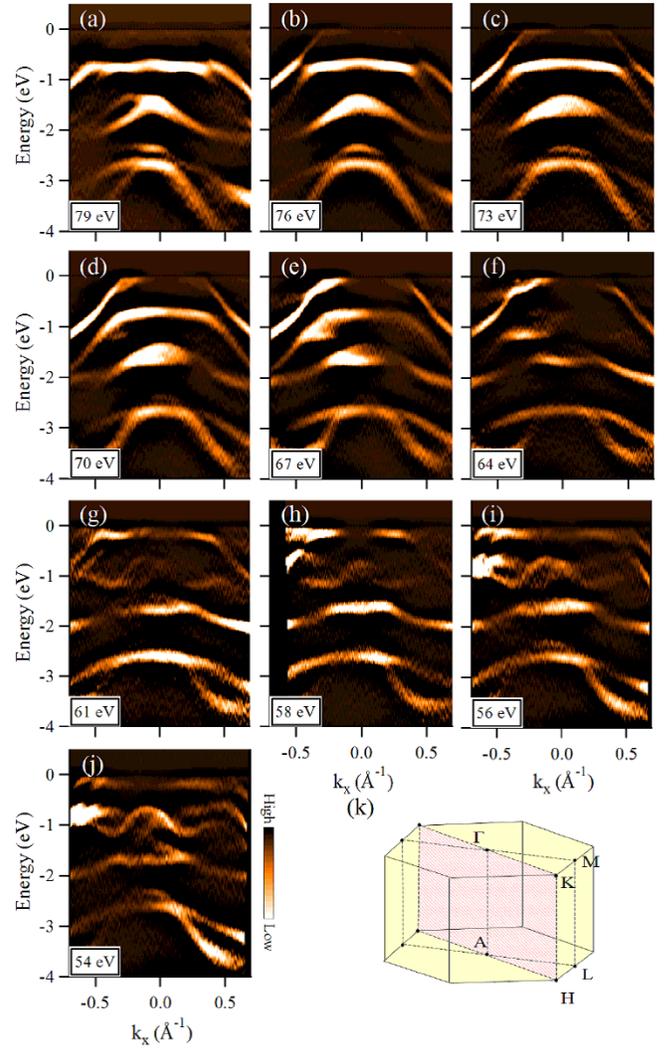}
\caption{(color online)
Second derivative plots of ARPES spectra approximately 
along the $\Gamma$-K or A-H direction
taken at $h\nu$ = 79 eV (a), 76 eV (b), 73 eV (c), 70 eV (d), 67 eV (e), 
64 eV (f), 61 eV (g), 58 eV (h), 56 eV (i), and 54 eV (j).
The dots indicate the Fermi surfaces predicted by the band-structure calculation.
}
\end{figure}

Figures 2(a)-(j) show the second derivative plots of ARPES spectra 
along the $\Gamma$-K or A-H direction taken at various photon energies
from $h\nu=79$ eV to $h\nu=54$ eV. 
Here, the momentum along the $\Gamma$-K or A-H direction 
of the Brillouin zone of IrTe$_2$ is defined as $k_x$. 
The momentum along the $\Gamma$-A direction is $k_z$, and $k_y$ is perpendicular 
to $k_x$ and $k_z$. The relationship between $k_x$, $k_y$, and $k_z$ is given by
$h\nu-W+V_0=\hbar^2/2m \times \sqrt{k_x^2+k_y^2+k_z^2}$ at photon energy of $h\nu$.
Here, the work function $W$ and $V_0$ are set to 4.4 eV, and 14 eV, respectively.
The momentum parallel to the surface is conserved at the electron emission
from the surface and is measured as $\sqrt{k_x^2+k_y^2} = 0.512 \times 
\sqrt{h\nu-W}\sin{\theta}$ (in unit of $\AA^{-1}$),
where $\theta$ is the emission angle from the surface normal direction.
Therefore, the momentum perpendicular to the surface is
given by $k_z = 0.512 \times \sqrt{(h\nu-W)\cos^2{\theta}+V_0}$,
and the photon energy $h\nu$ can be translated into $k_z$.

\begin{figure}
\includegraphics[width=7cm]{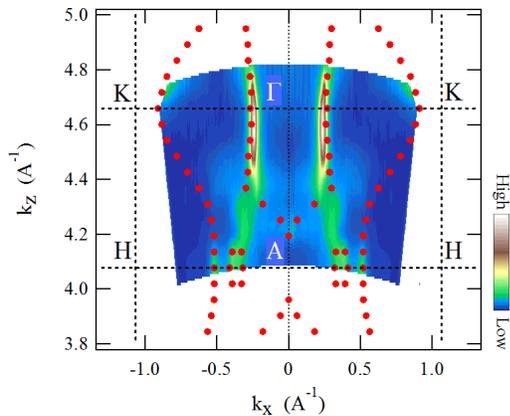}
\caption{(color online)
Fermi surface map in the $k_x$-$k_z$ plane.
Here, $k_x$ and $k_z$ are electron momenta along 
the x-direction (the $\Gamma$-K or A-H direction)
and the z-direction (the $\Gamma$-A direction), respectively.
}
\end{figure}

As seen in Figures 2(a)-(j), the band dispersions above -1.5 eV
strongly depend on the photon energy whereas those below -1.5 eV 
are relatively insensitive to the photon energy.
According to the band-structure calculation, the band dispersions above -1.5 eV 
are mainly constructed from the Ir 5$d$ $t_{2g}$ and the Te 5$p$ orbitals. 
Under the trigonal ligand field, the Ir 5$d$ $t_{2g}$ orbitals
are split into the Ir 5$d$ $a_{1g}$ and Ir 5$d$ $e'_{g}$ orbitals
whose lobes are directed to the out-of-plane ($z$-axis) direction 
and the in-plane ($xy$-plane) direction, respectively.
The Ir 5$d$ $a_{1g}$ orbital ($e'_{g}$ orbitals) strongly hybridize with 
the Te 5$p_z$ (5$p_x$ and 5$p_y$) orbitals which have the larger (smaller) 
5$p$-5$p$ transfer integrals between the IrTe$_2$ layers. Therefore, 
the Ir 5$d$ $a_{1g}$-Te 5$p_z$ bands have three dimensional
band dispersion while the Ir 5$d$ $e'_{g}$-Te 5$p_{x,y}$ bands
are expected to be two dimensional. This situation is roughly
obtained in the band structure calculation without the spin-orbit
interaction, and is not consistent with the ARPES result.
On the other hand, in the band structure calculation with
the spin-orbit interaction, the Ir 5$d$ $a_{1g}$ and $e'_{g}$ 
orbitals are heavily mixed due to the strong spin-orbit interaction
of the Ir 5$d$ orbitals, and all the Ir 5$d$ $t_{2g}$-Te 5$p$ 
bands exhibit large three dimensionality. The photon energy dependence 
of the observed band dispersions displayed in Fig. 2 shows that 
all the Ir 5$d$ $t_{2g}$-Te 5$p$ near $E_F$ strongly depend on $k_z$, 
consistent with the prediction of the band structure calculation 
considering the spin-orbit interaction.

\begin{figure}
\includegraphics[width=8.7cm]{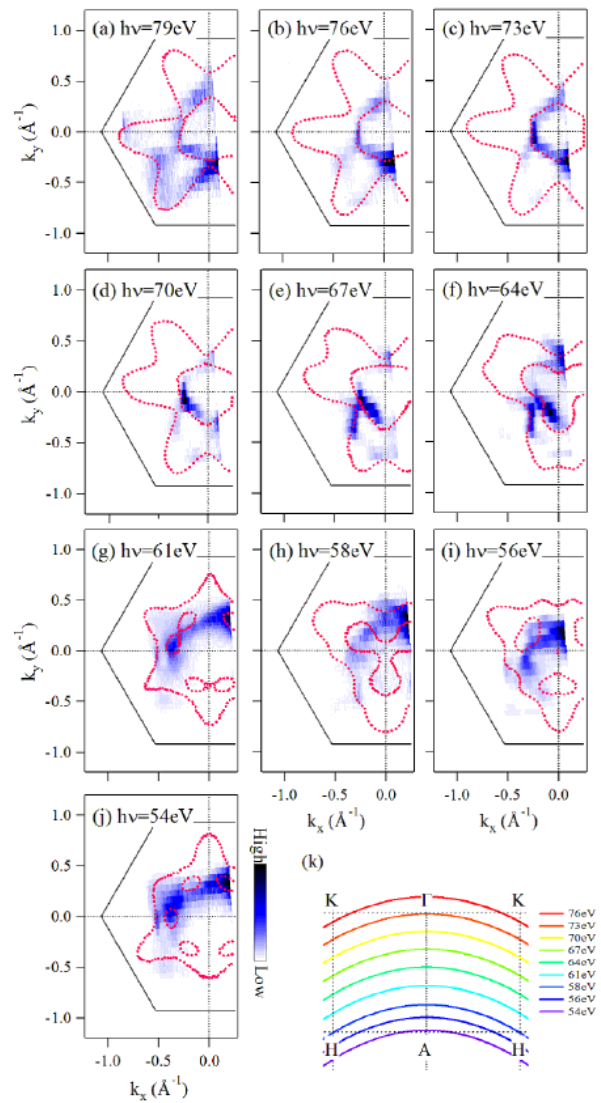}
\caption{(color online)
Fermi surface maps in the $k_x$-$k_y$ plane
with $k_z$ = 4.82 $\AA^{-1}$ (a), $k_z$ = 4.74 $\AA^{-1}$ (b), 
$k_z$ = 4.65 $\AA^{-1}$ (c), $k_z$ = 4.57 $\AA^{-1}$ (d), 
$k_z$ = 4.48 $\AA^{-1}$ (e), $k_z$ = 4.39 $\AA^{-1}$ (f), 
$k_z$ = 4.30 $\AA^{-1}$ (g), $k_z$ = 4.21 $\AA^{-1}$ (h), 
$k_z$ = 4.15 $\AA^{-1}$ (i), and $k_z$ = 4.08 $\AA^{-1}$ (j).
(k) Relationship between $k_x$ and $k_z$ for varions photon energies.
The Fermi surface maps are compared with the calculated
Fermi surfaces which are indicated by the solid curves.
}
\end{figure}

In Fig. 3, the ARPES intensity at $E_F$ is plotted
as a function of $k_x$ and $k_z$ for various photon energies
and is compared with the calculated Fermi surfaces indicated 
by the dots. The observed ARPES intensity at $E_F$
roughly follows the calculated Fermi surfaces.
In the region around 
the $\Gamma$ point ($4.8 \AA^{-1} < k_z < 4.4 \AA^{-1}$),
the observed Fermi surfaces are rather simple. 
The inner Fermi surface is almost parallel to the $\Gamma$-A axis, 
and the distance between the $\Gamma$-A axis and the inner Fermi surface
is $\sim$ 0.25 $\AA^{-1}$.
In the region around the A point ($4.4 \AA^{-1} < k_z < 4.1 \AA^{-1}$),
the inner and outer Fermi surfaces show the complicated structure.
In particular, whereas the calculation predicts that the inner Fermi 
surfaces are broken around $k_z \sim 4.2 \AA^{-1}$ in the $k_x$-$k_z$ plane,
substantial spectral weight is observed in this region due to the flat dispersion
[see Figs. 1(g)-(j)]. The residual spectral weight at $E_F$ due to 
the flat band dispersion would play important role in the nesting picture
although the nesting vector suggested by Yang {\it et al.} \cite{Yang2012}
is not seen in the Fermi surface map in the $k_x$-$k_z$ plane.

In Figs. 4(a)-(j), the ARPES intensities at $E_F$ are plotted
as functions of $k_x$ and $k_y$, and the Fermi surface maps 
in the $k_x$-$k_y$ plane are compared with the prediction of
the band structure calculations for various photon energies. 
Here, the relationship between $k_z$ and $k_x$ ($k_y$) in Fig. 4(k) 
is considered in the calculation. 
In going from Fig. 4(a) to Fig. 4(e), the inner and ourter Fermi 
surfaces are relatively insensitive to the $k_z$ value both 
in the calculation and the experiment. 
In the calculation, the inner Fermi surface is separated from 
the outer one along the $\Gamma$-M line in Figs. 4(b)-(d) and 
finally almost touches to the outer one in Figs. 4(a) and (e).
On the other hand, in the experiment, substantial spectral weight
is observed already in Figs. 4(b)-(d) in the region where the calculated
inner and outer Fermi surfaces touch to one another in Figs. 4(a) and (e).
This subtle discrepancy between the experiment 
and the theory suggests that the coupling between the inner and 
outer Fermi surfaces is more important
than the prediction of the band structure calculation.
In going from Fig. 4(f) to Fig. 4(i), the inner and outer 
Fermi surfaces dramatically change with $k_z$, and the experimental
result is basically consistent with the band structure calculation.
However, substantial spectral weight is observed in the region 
between the calculated inner and outer Fermi surfaces.

The intervening between the inner and outer Fermi surfaces 
can be related to the substantial distribution of Ir-Ir distance 
observed by the recent EXAFS study on IrTe$_2$ \cite{Joseph2013}. 
The Ir-Ir bond disorder indicates out-of-plane displacement 
of Ir atoms which can mix the $a_{1g}$ and $e_g$ orbitals 
in addition to the spin-orbit interaction. Most probably, 
the Ir-Ir bond disorder with the out-of-plane displacement
is driven by the Te-Te interaction between the layers 
\cite{Fang2012}, and therefore, the Te 5$p$ orbitals play
important roles both in the intervening between 
the Fermi surfaces and in the three dimensionality 
of the Fermi surfaces. Here, one can speculate that
the orbital fluctuations (and the superconductivity)
would be enhanced by the particular three-dimensional 
Fermi surfaces via the interlayer Te-Te interaction.

In conclusion, we have studied the three-dimensional multi-band Fermi surfaces
of Ir$_{0.95}$Pt$_{0.05}$Te$_2$ using ARPES and band structure calculation. 
The strong Te 5$p$-Te 5$p$ hybridization between the IrTe$_2$ triangular 
lattice layers is responsible for the strong $k_z$ dependence of the band 
dispersions which is confirmed by the ARPES experiment. 
The observed inner and outer Fermi surfaces with strong $k_z$ dependence 
are basically consistent with the band structure calculation. 
The strong Te-Te interaction between the layers is responsible 
for the three dimensionality of the Fermi surfaces.
The intervening between the inner and outer Fermi surfaces is larger 
in the ARPES result than the calculation, suggesting possibility 
of orbital fluctuation which would be enhanced by the Te-Te interaction
and the three-dimensional Fermi surfaces.

The authors would like to thank Profs. A. Damascelli and H.-J. Noh for valuable discussions.
This work was partially supported by Grants-in-Aid from the Japan Society of 
the Promotion of Science (JSPS) (22540363, 23740274, 24740238, 25400356) and 
the Funding Program for 
World-Leading Innovative R\&D on Science and Technology (FIRST Program) from JSPS.
T.T. and D.O. acknowledge supports from the JSPS Research Fellowship for Young Scientists. 
The synchrotron radiation experiment was performed with the approval of 
Photon Factory, KEK (2013G021).

\end{document}